\documentclass[aps,prl,twocolumn,groupedaddress,showpacs]{revtex4-1} 
\usepackage{graphicx}
\usepackage{color}

\begin{document}

 
\title{
Work Output and Efficiency at Maximum Power of Linear Irreversible Heat Engines Operating with a  
Finite-Sized Heat Source
} 


\author{Yuki Izumida$^1$ and Koji Okuda$^2$}
\affiliation{$^1$Department of Information Sciences, 
Ochanomizu University, Tokyo 112-8610, Japan\\
$^2$Division of Physics, Hokkaido University, Sapporo 060-0810, Japan}



\begin{abstract}
We formulate the work output and efficiency 
for linear irreversible heat engines working between a finite-sized hot heat source and an infinite-sized cold heat reservoir until the total system reaches the final thermal equilibrium state with a uniform temperature.
We prove that when the heat engines operate at the maximum power under the tight-coupling condition without heat leakage the work output is just half of the exergy, 
which is known as the maximum available work extracted from a heat source.
As a consequence, the corresponding efficiency is also half of its quasistatic counterpart.
\end{abstract} 


\maketitle

{\sl Introduction.--} 
Utilizing heat as the motive power has been an indispensable basis for our modern industrial society.
The Carnot theorem is a cornerstone for heat-energy conversion, where a heat engine converts the heat $Q_h$ into the work output $W$ between the hot heat reservoir
(source) at the temperature $T_h$ and the cold heat reservoir (sink) at the temperature $T_c$ ($T_c < T_h$).
The theorem states that the efficiency $\eta=\frac{W}{Q_h}$ of the heat-energy conversion 
is bounded from above as 
\begin{eqnarray}
\eta \le 1-\frac{T_c}{T_h}\equiv \eta_{\rm C} \ ({\rm Carnot \ efficiency}),\label{eq.carnot_theorem} 
\end{eqnarray}
showing that
we need to discard a certain amount of heat $Q_c\equiv Q_h-W$ into the cold heat reservoir.
Although the upper bound is achieved by reversible heat engines with infinitely slow (quasistatic) operation such as in the Carnot cycle, 
the power (work output per unit time), which is another important performance criterion, vanishes
in the quasistatic limit by definition. 
In this regard, for more practical relevance, the efficiency at maximum power $\eta^*$ has 
been intensively studied~\cite{CA,Y,N,LL,G,B,SNGAL,VB,CH,ELB2,SS,GMS,EKLB2,US,IO4}.

In terms of practicality, it would also be relevant to consider the finiteness of the heat source.
Even though the hot heat source may usually be treated as an infinite-sized heat reservoir in the thermodynamics of heat engines, it can actually be ``fuel" as a finite resource with a finite amount of substance (e.g., burning coal in a steam engine) but not necessarily an infinite resource like the heat reservoir. 
This issue is becoming increasingly important due to the need for an urgent solution to the worldwide depletion of energy resources. To achieve a sustainable society, we need to consider this kind of fundamental thermodynamic limitation.

Reversible heat engines utilizing a finite-sized hot heat source have been considered in the context of exergy~\cite{JWG,Z,C,MSBB}.
Exergy is the maximum available work defined 
as the upper bound of the work output extracted from a finite-sized hot heat source 
until the hot heat source is brought to the final thermal equilibrium state sharing the same temperature as a cold heat reservoir 
(see the details below). 
The corresponding efficiency also achieves a maximum, 
but again, the power vanishes while the maximum work is extracted.
Even though some previous literature addressed the available work extracted from a finite-sized heat source during finite-time operation and the corresponding efficiency~\cite{ARB,ABOS,YC,ORB}, 
their results depended on phenomenological models with specific assumptions. 

Our purpose is to formulate the work output and efficiency at the maximum power 
for linear irreversible heat engines working between a finite-sized hot heat source and an infinite-sized cold heat reservoir until the total system reaches the final thermal equilibrium state with a uniform temperature.
Our theory is based on a general framework of linear irreversible thermodynamics~\cite{O,GM}. 
We prove that, when the heat engines operate at the maximum power under the tight-coupling condition without heat leakage, the work output is just half of the exergy, and as a consequence, the corresponding efficiency is also half of its quasistatic counterpart.
Our results also include $\eta^*=\frac{\eta_{\rm C}}{2}$, obtained in~\cite{VB} for linear irreversible heat engines working between two infinite-sized heat reservoirs under the tight-coupling condition as a special case.

{\sl Exergy.--} 
\begin{figure}[t]
\includegraphics[scale=0.27]{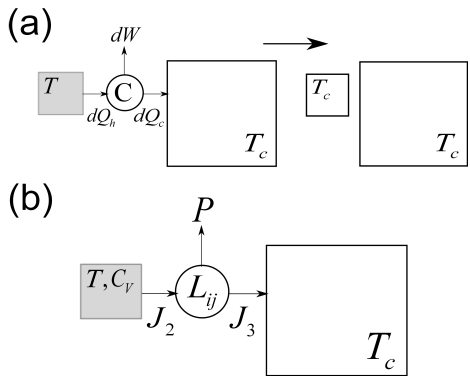}
\caption{
(a) Energetically infinitesimal 
Carnot cycle working between a finite-sized hot heat source at a temperature $T$ 
and the cold heat reservoir at the temperature $T_c$.
The total system ends up in the final thermal equilibrium state sharing the uniform temperature $T_c$.  
(b) Linear irreversible heat engine working between a finite-sized hot heat source at $T$ with a heat capacity at constant volume $C_V$ and a cold heat reservoir at $T_c$.}\label{fig}
\end{figure}
The maximum work can be extracted by an energetically infinitesimal reversible heat engine between the temperatures $T$ and $T_c$ (e.g., the Carnot cycle), where $T$ is the temperature of the hot heat source. 
Such a heat engine successively transforms the infinitesimal heat $dQ_h$ into the infinitesimal work $dW$ 
with the efficiency $\eta^T$ at each $T$ being the highest Carnot efficiency $\eta_{\rm C}^{T}\equiv1-\frac{T_c}{T}$
until the temperature $T$ of the hot heat source decreases from $T_h$ to $T_c$
[Fig.~\ref{fig} (a)]. 
The total work extracted by a heat engine from the finite-sized hot heat source 
defined by $W=\int dW=\int \eta^T dQ_h$ is bounded from above for the scenario in Fig.~\ref{fig} (a) as 
\begin{eqnarray}
W \le \int \eta_{\rm C}^TdQ_h&&=-\int_{T_h}^{T_c} \eta_{\rm C}^TC_VdT\nonumber\\
&&=U_h-U_c-T_c(S_h-S_c)\equiv E,\label{eq.def_e}
\end{eqnarray} 
where $U_i$ ($S_i$) ($i=h, c$) are the internal energies (entropies) of the initial and final equilibrium states of the hot heat source 
using the definitions $U_h-U_c\equiv \int_{T_c}^{T_h} C_VdT$ and $S_h-S_c\equiv \int_{T_c}^{T_h} \frac{C_V}{T}dT$, with $C_V=C_V(T)$ denoting the heat capacity at constant volume of the hot heat source.
This maximum work $E$ is called the ``exergy"~\cite{Z}.
Although the significance of the exergy is rarely considered in a context of physics (but see, e.g.,~\cite{C,DL}), 
it is a practically important concept that is often introduced in thermodynamics textbooks for engineers~\cite{MSBB}.
The exergy is not a state function since it depends on a condition under which the hot heat source supplies heat to the heat engine during the process. Although here we consider a constant-volume condition for simplicity, the concept of the exergy itself can also be considered for other thermodynamic conditions (e.g., a constant-pressure condition).
The corresponding efficiency $\eta=\frac{W}{Q_h}=\frac{W}{U_h-U_c}$ is bounded from above as~\cite{OAMB,L} 
\begin{eqnarray}
\eta \le \frac{E}{U_h-U_c}=1-\frac{T_c(S_h-S_c)}{U_h-U_c}\equiv \eta_{\rm max}\label{eq.effi_qs_thermo}.\label{eq.def_effi_qs}
\end{eqnarray}
In the case that the hot heat source is a heat reservoir as $C_V \to \infty$, 
$\eta_{\rm max}$ recovers the usual Carnot efficiency $\eta_{\rm C}$ by the definition $\frac{U_h-U_c}{T_h}=\frac{Q_h}{T_h}=S_h-S_c$ for an isothermal environment.
Consequently, Eq.~(\ref{eq.def_effi_qs}) for a finite-sized hot heat source is a generalization of the Carnot theorem in Eq.~(\ref{eq.carnot_theorem}) for an infinite-sized hot heat source.

{\sl Linear irreversible thermodynamics framework.--} 
The setup of our model is as follows [Fig.~\ref{fig} (b)]: consider a linear irreversible heat engine~\cite{VB} 
working between a finite-sized hot heat source and a cold heat reservoir at the temperature $T_c$. 
We assume that the hot heat source is always in an equilibrium state with a well-defined 
temperature $T$ between $T_c$ and $T_h$ 
and the heat capacity at constant volume $C_V$.
Initially at time $t=0$, $T$ is assumed to be $T_h$. 
We also assume that the heat engine successively 
transforms the heat (a small part of the internal energy of the hot heat source) into work at an efficiency 
depending on the working regime, and $T$ continuously decreases from $T_h$ to $T_c$ accompanying the operation of the heat engine.
We note that these assumptions could physically be realized under the two following relevant conditions: 
(i) the hot heat source rapidly relaxes to an equilibrium state with a well-defined temperature 
upon providing the heat energy to the heat engine and
(ii) the internal energy of the hot heat source 
is sufficiently larger than the amount of the heat energy supplied to the heat engine per cycle or unit time, which implies that 
it takes sufficiently many cycles or a long time to deplete the hot heat source.

The behavior of our heat engine during the process is described by linear irreversible thermodynamics, 
starting from consideration of 
the entropy production rate $\dot{\sigma}$ of the total system consisting of the heat engine, hot heat source, and cold heat reservoir. 
Here, the dot indicates a quantity per unit time or the derivative with respect to time.
Because the internal state of the heat engine returns to the original state after a unit time for cyclic heat engines if the unit time is chosen as one cycle or remains unchanged for steady-state heat engines, we express $\dot{\sigma}$ only by the sum of the entropy increase rates of the hot heat source and cold heat reservoir as  
\begin{eqnarray}
\dot{\sigma}=-\frac{\dot{Q}_h}{T}+\frac{\dot{Q}_c}{T_c}=
-\frac{\dot{W}}{T_c}+\dot{Q}_h\biggl(\frac{1}{T_c}-\frac{1}{T}\biggr).
\end{eqnarray}
Defining a generalized external force $F$ that acts on its conjugate variable $x$, we write
the power $\dot{W}$ as $\dot{W}=-F\dot{x}$.
Then, we naturally decompose $\dot{\sigma}=\frac{F\dot{x}}{T_c}+\dot{Q}_h\bigl(\frac{1}{T_c}-\frac{1}{T}\bigr)$ as
$\dot{\sigma}=J_1X_1+J_2X_2$, 
defining the thermodynamic fluxes $J_1$ (the motion speed of the heat engine) and $J_2$ (the heat flux from the hot heat source)
in response to their conjugate thermodynamic forces $X_1$ (the external force) and $X_2$ (the thermal gradient), respectively, as
\begin{eqnarray}
&&X_1\equiv \frac{F}{T_c}, \ \ J_1\equiv \dot{x},\\
&&X_2\equiv \frac{1}{T_c}-\frac{1}{T}=\frac{T-T_c}{TT_c}, \ \ J_2\equiv \dot{Q}_h.
\end{eqnarray} 
We assume that these variables are related to the linear Onsager relations~\cite{VB,O,GM} as
\begin{eqnarray}
&&J_1=L_{11}X_1+L_{12}X_2,\label{eq.J1}\\ 
&&J_2=L_{21}X_1+L_{22}X_2,\label{eq.J2} 
\end{eqnarray} 
where $L_{ij}$ denotes the Onsager coefficients with the reciprocal relation $L_{12}=L_{21}$.
We note that, from the nonnegativity of $\dot{\sigma}=J_1X_1+J_2X_2\ge 0$, the possible values of $L_{ij}$ are restricted as 
\begin{eqnarray}
L_{11}\ge 0, \ L_{22}\ge 0, \ L_{11}L_{22}-L_{12}L_{21}\ge 0.\label{eq.onsager_coeffi_range}
\end{eqnarray}
For a simpler formulation, we change the variable from $X_1$ to $J_1$ as $X_1=(J_1-L_{12}X_2)/L_{11}$ through Eq.~(\ref{eq.J1}) and 
replace the Onsager relations with $J_2$ and the heat flux to the cold heat reservoir $J_3\equiv \dot{Q}_c=\dot{Q}_h-\dot{W}=J_2+J_1 X_1 T_c$~\cite{IO4}:
\begin{eqnarray} 
&&J_2=\frac{L_{21}}{L_{11}}J_1+L_{22}(1-q^2)X_2,\label{eq.J2-2}\\  
&&J_3=\frac{L_{21}T_c}{L_{11}T}J_1+L_{22}(1-q^2)X_2+\frac{T_c}{L_{11}} {J_1}^2,\label{eq.J3-2}  
\end{eqnarray} 
where $q$ is the coefficient of the coupling strength defined as~\cite{KC}
\begin{eqnarray}
q\equiv \frac{L_{21}}{\sqrt{L_{11}L_{22}}}, \ \ (|q|\le 1).
\end{eqnarray}  
The restriction of $q$ comes from Eq.~(\ref{eq.onsager_coeffi_range}).
Each term in Eqs.~(\ref{eq.J2-2}) and (\ref{eq.J3-2}) has a clear physical interpretation~\cite{IO4}:
the first terms in Eqs.~(\ref{eq.J2-2}) and (\ref{eq.J3-2}) imply the reversible heat transfer from the hot heat source and that to the cold heat reservoir, which do not contribute to $\dot{\sigma}$ at all. 
The second terms in Eqs.~(\ref{eq.J2-2}) and (\ref{eq.J3-2}) imply the heat leakage from the hot heat source to the cold heat reservoir 
with $\frac{L_{22}(1-q^2)}{TT_c}$ being the thermal conductance,
which remain irrespective of the operation of the heat engine with $|q|\ne 1$. 
The third term in Eq.~(\ref{eq.J3-2}) implies the dissipation into the cold heat reservoir accompanying the operation of the heat engine.

Under the condition $|q|=1$ called ``tight-coupling," 
the heat-leakage terms vanish, 
and the heat fluxes go to zero simultaneously in the quasistatic limit of $J_1\to 0$. 
This idealized but most interesting condition can be realized, e.g., by a finite-time Carnot cycle~\cite{IO2}.

The power $\dot{W}=J_2-J_3$ is rewritten by using Eqs.~(\ref{eq.J2-2}) and (\ref{eq.J3-2}) as 
\begin{eqnarray}
\dot{W}=\frac{L_{12}}{L_{11}}\eta_{\rm C}^T J_1-\frac{T_c}{L_{11}}J_1^2.\label{eq.power}
\end{eqnarray}

The decreasing rate of the temperature $T$ of the hot heat source during the operation of the heat engine is given by
\begin{eqnarray}
C_V\frac{dT}{dt}=C_V\dot{T}=-J_2,\label{eq.T_change}
\end{eqnarray}
where $J_2$ is given by Eq.~(\ref{eq.J2-2}).
Equation (\ref{eq.T_change}) can also be considered as a relation that connects the temperature $T$ to the time $t$.
Although $J_i$, $X_i$, $L_{ij}$, and $C_V$
may depend on $T$ [or $t$ through Eq.~(\ref{eq.T_change})] in general, we write the $T$ dependence explicitly only when we stress it. Equations (\ref{eq.J2-2}) and (\ref{eq.J3-2}) together with Eq.~(\ref{eq.T_change}) constitute the 
time-evolution equations of $T(t)$ for a given working regime $J_1(t)$.

After these preparations, we consider the work output and efficiency of this system.
The heat from the hot heat source $Q_h$ and the work output $W$ 
between the initial time $t=0$ and the final time $t=\tau$
are given as 
\begin{eqnarray}
Q_h&&=\int_0^{\tau} J_2(t)dt=-\int_{T_h}^{T_c} C_VdT=U_h-U_c,\label{eq.total_heat}\\
W&&=\int_0^{\tau} \dot{W}(t)dt
=U_h-U_c-\int_0^{\tau}J_3(t)dt,\label{eq.total_work}
\end{eqnarray}
respectively, where we used Eq.~(\ref{eq.T_change}), $T(0)=T_h$, and $T(\tau)=T_c$.
$W$ is a functional of $J_1$ [or equivalently $T$ through Eq.~(\ref{eq.T_change})] via $J_3$ in Eq.~(\ref{eq.J3-2}).
We also express the total power $P$ and the efficiency $\eta$ as
\begin{eqnarray}
P&&=\frac{W}{\tau}=\frac{U_h-U_c-\int_0^{\tau}J_3(t)dt}{\tau},\label{eq.total_power}\\
\eta&&=\frac{W}{Q_h}=1-\frac{\int_0^{\tau}J_3(t)dt}{U_h-U_c},\label{eq.total_efficiency}
\end{eqnarray}
respectively. 

{\sl Main results.--}
We maximize Eq.~(\ref{eq.total_power}) by first minimizing the integral $\int_0^{\tau}J_3(t)dt$ in Eq.~(\ref{eq.total_power}) under the fixed time $\tau$ and then maximizing $P(\tau)$ as follows.
By solving Eq.~(\ref{eq.T_change}) with respect to $J_1$, we express $J_1(t)$ as a function of $T$ and $\dot{T}$ as
\begin{eqnarray}
J_1(T, \dot{T})=-\frac{L_{11}}{L_{21}}C_V\dot{T}-\frac{L_{11}L_{22}(1-q^2)}{L_{21}}X_2.
\end{eqnarray}
Then we express the integrand $J_3(t)=J_3(T, \dot{T})$ as
\begin{eqnarray}
J_3(T,\dot{T})&&=-T_cC_V\frac{\dot{T}}{T}+\frac{L_{22}(1-q^2)X_2^2T_c}{q^2}+\frac{T_cC_V^2}{q^2L_{22}}\dot{T}^2\nonumber \\
&& +\frac{2T_c}{q^2}C_V\dot{T}(1-q^2)X_2\label{eq.variation_eq}
\end{eqnarray}
by substituting $J_1(T, \dot{T})$ into Eq.~(\ref{eq.J3-2}).
We minimize the integral $\int_0^{\tau}J_3(T,\dot{T})dt$ in Eq.~(\ref{eq.total_power})
by solving the following Euler-Lagrange (EL) equation for $T(t)$:
\begin{eqnarray}
\frac{d}{dt}\left(\frac{\partial J_3(T,\dot{T})}{\partial \dot{T}}\right)-\frac{\partial J_3(T,\dot{T})}{\partial T}=0.\label{eq.el_eq}
\end{eqnarray}
By substituting Eq.~(\ref{eq.variation_eq}) into Eq.~(\ref{eq.el_eq}), we obtain the EL equation to be solved as
\begin{eqnarray}
\frac{2C_V^2}{q^2L_{22}}\ddot{T}+\dot{T}^2\frac{\partial}{\partial T}\Bigl(\frac{C_V^2}{q^2L_{22}}\Bigr)\nonumber\\
-\frac{\partial}{\partial T}\Bigl(\frac{L_{22}(1-q^2)X_2^2}{q^2}\Bigr)=0.\label{eq.el_simp}
\end{eqnarray}
Note that $C_V$, $L_{22}$, and $q$ may depend on $T$.
By multiplying both sides by $\dot{T}$, Eq.~(\ref{eq.el_simp}) is simplified to
\begin{eqnarray}
\frac{d}{dt}\Biggl(\frac{C_V^2}{q^2L_{22}}\dot{T}^2-\frac{L_{22}(1-q^2)X_2^2}{q^2}\Biggr)=0.\label{eq.el_eq_solution}
\end{eqnarray}
It may be difficult to find an explicit form of the general solution of $T(t)$ from Eq.~(\ref{eq.el_eq_solution}).

However, under the tight-coupling condition $|q|=1$ which is idealized but is the most interesting condition, we can calculate the integral 
$\int_0^{\tau}J_3(T,\dot{T})dt$ by utilizing Eq.~(\ref{eq.el_eq_solution}) without solving it explicitly as follows.
By integrating Eq.~(\ref{eq.el_eq_solution}), we obtain
\begin{eqnarray}
\frac{C_V}{\sqrt{L_{22}}}\dot{T}=A \ ({\rm integral \ const}).\label{eq.el_eq_solution2}
\end{eqnarray}
By integrating Eq.~(\ref{eq.el_eq_solution2}) from $t=0$ to $t=\tau$, we obtain
\begin{eqnarray}
A=\frac{\int_{T_h}^{T_c}\frac{C_V}{\sqrt{L_{22}}}dT}{\tau}\equiv \frac{B}{\tau},\label{eq.integral_const}
\end{eqnarray}
where $B$ is a constant independent of $\tau$. By using Eqs.~(\ref{eq.variation_eq}), (\ref{eq.el_eq_solution2}), and (\ref{eq.integral_const}), we calculate the integral $\int_0^{\tau}J_3(T,\dot{T})dt$ and the total power in Eq.~(\ref{eq.total_power}) as
\begin{eqnarray}
\int_0^{\tau}J_3dt&&=-T_c \int_0^{\tau} \left(\frac{C_V \dot{T}}{T}-A^2\right) dt\nonumber \\
&&=T_c(S_h-S_c)+\frac{T_cB^2}{\tau},\label{eq.integrand_after_minimization}\\
P_{|q|=1}&&=\frac{E}{\tau}-\frac{T_cB^2}{\tau^2},\label{eq.total_power_after_variation}
\end{eqnarray}
respectively. 
Note that $\eta_{|q|=1}$ in Eq.~(\ref{eq.total_efficiency}) $\to \eta_{\rm max}$ in Eq.~(\ref{eq.def_effi_qs}) and $P_{|q|=1}\to 0$ in the quasistatic limit $\tau \to \infty$.

By maximizing Eq.~(\ref{eq.total_power_after_variation}) with respect to $\tau$, we obtain the maximum power as
\begin{eqnarray}
P_{|q|=1}^*=\frac{E^2}{4T_cB^2}\ \  \left({\rm at}\ {\tau}^*=\frac{2T_cB^2}{E}\right).\label{eq.total_time_at_pmax}
\end{eqnarray}
Using Eq.~(\ref{eq.total_time_at_pmax}), 
we finally obtain the work output at the maximum power under the tight-coupling condition as
\begin{eqnarray}
W^*_{|q|=1}=P_{|q|=1}^{*}\tau^*=\frac{E}{2},\label{eq.work_global_pmax}
\end{eqnarray}
which is just half of the exergy.
Then, as a consequence of Eq.~(\ref{eq.work_global_pmax}), we also conclude that the efficiency at the maximum power 
under the tight-coupling condition $\eta^*_{|q|=1}$ is also half of the maximum efficiency in Eq.~(\ref{eq.def_effi_qs}) as
\begin{eqnarray}
\eta^*_{|q|=1}=\frac{W^*_{|q|=1}}{U_h-U_c}=\frac{1}{2}\eta_{\rm max}.\label{eq.effi_at_max_pow}
\end{eqnarray}
We note that in the case where the hot heat source is a hot heat reservoir as $C_V \to \infty$, 
$\eta^*_{|q|=1}$ in Eq.~(\ref{eq.effi_at_max_pow}) recovers $\frac{\eta_{\rm C}}{2}$, as previously derived for the linear irreversible heat engines working between the heat reservoirs~\cite{VB}.
Although we considered only the tight-coupling case,  
it is quite natural to expect that Eqs.~(\ref{eq.work_global_pmax}) and (\ref{eq.effi_at_max_pow}) also serve as the {\it upper bounds} for the work output and efficiency at the maximum power, respectively,   
as the heat-leakage terms in Eqs.~(\ref{eq.J2-2}) and (\ref{eq.J3-2}) do not contribute to the work output at all 
but just reduce the available internal energy of the hot heat source. 
This would also be consistent with the fact that the tight-coupling case 
serves as the upper bound of the efficiency at the maximum power $\eta^*$ for linear irreversible heat engines with the infinite-sized heat reservoir $C_V \to \infty$ 
as in Eq.~(9) of~\cite{VB} because our theory should also lead to Eq.~(9) of~\cite{VB} in that limit.

{\sl Concluding remarks.--} 
We formulated the work output and efficiency for linear irreversible heat engines working between a finite-sized hot heat source and an infinite-sized cold heat reservoir until the total system reaches the final thermal equilibrium state with a uniform temperature.
We proved that, when the heat engines operate at the maximum power under the tight-coupling condition without heat leakage, 
the work output is just half of the exergy reached at the quasistatic limit, and
as a consequence, the corresponding efficiency is also half of its quasistatic counterpart.
Because our results can be applied to any type of working substance and do not assume any  
specific form of the Onsager coefficients and the heat capacity, 
we expect that they are universal, as are Eqs.~(\ref{eq.def_e}) and (\ref{eq.def_effi_qs}) derived by 
equilibrium thermodynamics.

Finally, we remark on some possible extensions of the present study.
First, further extension of our formulation to systems operating with a finite-sized cold heat sink instead of a cold heat reservoir~\cite{ORB,L,YC} is straightforward. Second, the idea presented here 
could naturally be applied to other types of heat devices such as refrigerators and heat pumps~\cite{CH2,BK,BB}
because their original function is cooling or heating a system to a desirable temperature~\cite{BK,BB}, 
where the system should be regarded as a finite-sized heat source or sink.
Third, an extension to nonlinear response regimes~\cite{ELB2,SS,GMS,EKLB2,US,IO4} would also be a challenging task for possible future studies.
Fourth, seeing that optimization of fluctuating nano- and micron-sized heat engines has become 
a hot topic with recent technological advances~\cite{BB2,ARJDSSL,RASSL}, we are naturally motivated to 
apply the exergetic concept to these systems. 
In relation to this, it is also interesting that a nonequilibrium equality such as the Crooks fluctuation theorem~\cite{Cr} has also been extended to a finite-sized heat reservoir~\cite{CTH}.
Fifth, we stress that our theory could serve as a ``zeroth approximation" for a possible future theory describing heat engines with a far-from-equilibrium heat source, such as a steam engine powered by burning coal. 

We expect that the present work not only expands the scope of finite-time thermodynamics
but also provides a design principle and an operational scheme for actual thermodynamic devices and power plants for a sustainable society.
\begin{acknowledgements} 
Y. I. acknowledges the financial support from a Grant-in-Aid for JSPS Fellows (Grant No. 25-9748).
The authors thank A. Calvo Hern\'{a}ndez for useful comments on the manuscript.
\end{acknowledgements} 

\end{document}